\documentclass[sigconf]{acmart}

\usepackage{amsmath,amssymb,amsfonts}
\usepackage{algorithmic}
\usepackage{graphicx}
\usepackage{textcomp}
\usepackage{booktabs}
\usepackage{multirow}
\usepackage{bm}
\usepackage[justification=centering]{caption}
\usepackage{balance}
\usepackage{float}
\usepackage{enumerate}
\usepackage{tcolorbox}
\usepackage{fbox}

\newtcolorbox{qoutebox}[3][]
{
  colframe=black!30!white,
  colback  = #2!10,
  #1,
}

\AtBeginDocument{%
  \providecommand\BibTeX{{%
    \normalfont B\kern-0.5em{\scshape i\kern-0.25em b}\kern-0.8em\TeX}}}

\copyrightyear{2022}
\acmYear{2022}
\setcopyright{acmcopyright}\acmConference[ICPC '22]{30th International Conference on Program Comprehension}{May 16--17, 2022}{Virtual Event, USA} \acmBooktitle{30th International Conference on Program Comprehension (ICPC '22), May 16--17, 2022, Virtual Event, USA}
\acmPrice{15.00}
\acmDOI{10.1145/3524610.3527884}
\acmISBN{978-1-4503-9298-3/22/05}

\begin{document}

\title[Understanding Code Snippets in Code Reviews]{Understanding Code Snippets in Code Reviews: A Preliminary Study of the OpenStack Community}


\author{Liming Fu}
\affiliation{%
   \institution{School of Computer Science \\Wuhan University}
   \city{Wuhan}
   \country{China}}
 \email{limingfu@whu.edu.cn}

\author{Peng Liang}
\affiliation{%
   \institution{School of Computer Science \\Wuhan University}
   \city{Wuhan}
   \country{China}}
 \email{liangp@whu.edu.cn}
 \authornote{This work was funded by the National Key R\&D Program of China with No. 2018YFB1402800 and the NSFC with No. 62172311.}

\author{Beiqi Zhang}
\affiliation{%
   \institution{School of Computer Science \\Wuhan University}
   \city{Wuhan}
   \country{China}}
 \email{zhangbeiqi@whu.edu.cn}

\renewcommand{\shortauthors}{L. Fu et al.}

\begin{abstract}
Code review is a mature practice for software quality assurance in software development with which reviewers check the code that has been committed by developers, and verify the quality of code. During the code review discussions, reviewers and developers might use code snippets to provide necessary information (e.g., suggestions or explanations). However, little is known about the intentions and impacts of code snippets in code reviews. To this end, we conducted a preliminary study to investigate the nature of code snippets and their purposes in code reviews. We manually collected and checked 10,790 review comments from the Nova and Neutron projects of the OpenStack community, and finally obtained 626 review comments that contain code snippets for further analysis. The results show that: (1) code snippets are not prevalently used in code reviews, and most of the code snippets are provided by reviewers. (2) We identified two high-level purposes of code snippets provided by reviewers (i.e., Suggestion and Citation) with six detailed purposes, among which, Improving Code Implementation is the most common purpose. (3) For the code snippets in code reviews with the aim of suggestion, around 68.1\% was accepted by developers. The results highlight promising research directions on using code snippets in code reviews. 
\end{abstract}


\maketitle

\section{Introduction}
Code review is widely known as one of the best practices in software development with the aim of verifying the source code quality. In a typical code review process, reviewers and developers work collaboratively through an asynchronous online discussions with review comments to ensure that the proposed code changes are of sufficiently high quality \cite{infoInCr}. During the discussions, reviewers and developers may exchange information with each other towards the issues to be addressed. In this context, code snippets, as a special form of code artifacts, might be used in review comments (e.g., reviewers might provide constructive suggestions with code snippets to help improving the quality of code). 

A recent study has been conducted to study the practice of link sharing and their intentions in code reviews, and to explore what information could be provided through link sharing \cite{Wang2021links}. Similar to links, code snippets are also considered as one of the measures to convey necessary information during the code review process. However, the natures of links and code snippets lead to different intentions and impacts in code reviews. To the best of our knowledge, little is known about the code snippets in code reviews. To \textbf{bridge this gap}, we conducted a preliminary study to provide a comprehensive overview of code snippets about its distribution, purposes, and acceptance in code reviews. We collected and analyzed review comments from two projects, Nova and Neutron, of the OpenStack community. The contribution of this work is twofold: (1) a study focusing on exploring code snippets in code reviews, and (2) a taxonomy of six purposes of providing code snippets in code reviews. Moreover, we made the dataset used in this work publicly available for replication purposes~\cite{replication-package}.

The remaining of this paper is structured as follows: Section \ref{related-work} surveys the related work on code reviews and code snippets, and Section \ref{Research Design} describes the research design of this study. The results of our study are presented in Section \ref{Results}, followed by the implications in Section \ref{Implications}. Section \ref{Threats to validity} clarifies the threats to the validity. Section \ref{Conclusions} concludes this work with future directions.

\section{Related Work} \label{related-work}
Many studies have explored the topic of information needs in code reviews. Pascarella \textit{et al.} investigated the information that reviewers need to conduct a proper code review \cite{infoInCr}. They identified the presence of seven high-level reviewers' information needs. Wang \textit{et al.} investigated link sharing and their intentions in code reviews \cite{Wang2021links}. They identified seven intentions behind link sharing in code reviews, in which providing context and elaborating are the most common intentions. Several other studies have focused on identifying various information about code snippets. Subramanian \textit{et al.} used code snippet analysis to extract their structural information in order to effectively identify API usage in the snippets \cite{Subramanian2013snippets}. Chatterjee \textit{et al.} also mined the natural language text associated with code snippets in software-related documents, e.g., API documentation and code reviews \cite{Chatterjee2017infoInCs}. We also identified symptoms of architecture erosion in code reviews~\cite{li2022sae}. Compared to the existing work that focuses on the information needs in code reviews and identifying information of code snippets, our work intends to explore the distribution, purposes, and acceptance of code snippets in code reviews.

\section{Research Design}\label{Research Design}
\subsection{Research Questions} \label{Research Questions}
The goal of this work formulated through the Goal Question Metric approach \cite{caldiera1994goal} is to investigate the existence of code snippets in code reviews \textbf{for the purpose of} exploration \textbf{with respect to} the distribution, purposes, and acceptance of code snippets \textbf{from the point of} developers and code reviewers \textbf{in the context of} OSS development. To achieve this goal, we formulated three Research Questions (RQs) as follows:

\noindent\textbf{RQ1. How many review comments contain code snippets in code reviews?}

\noindent\textbf{RQ2. What are the purposes of code snippets provided by reviewers in code reviews?}

\noindent\textbf{RQ3. How many review comments that contain code snippets are accepted by developers?}

\subsection{Data Collection}
This study focuses on investigating the code snippets in code reviews collected from the OpenStack\footnote{\url{https://www.openstack.org/}} community. OpenStack is a set of software tools for building and managing cloud computing platforms, and is supported by many large companies. We deemed OpenStack to be appropriate for this study because it has made a big investment in code reviews for several years \cite{HiraoCode} and is widely used in many studies related to code reviews (e.g., \cite{HiraoCode, hanCodeSmell}). The OpenStack community is composed of several projects, and we selected two of the most active projects as our investigated projects (based on the number of closed code changes), i.e., Nova\footnote{\url{https://github.com/openstack/nova}} and Neutron\footnote{\url{https://github.com/openstack/neutron}}.

The code review process of OpenStack is supported by Gerrit\footnote{\url{https://www.gerritcodereview.com/}}, a web-based code review platform built on the top of Git. By using the RESTful API provided by Gerrit, we were able to collect all the closed code changes of the two selected projects that were updated in 2020. We then extracted all available review comments of these code changes and stored the data in a local file for further analysis. In total, we collected a dataset that contains 3,447 code changes and 20,978 review comments.

\subsection{Data Labelling, Extraction and Analysis}
Based on the recent data, OpenStack projects contain around 13 million lines of code, and most of them are written in Python\footnote{\url{https://www.openhub.net/p/openstack}}. Considering this, we decided to only focus on the review comments that contain Python code snippets. Specifically, the \textbf{inclusion criterion} is that the review comments must include a portion of source code (or pseudocode), which contains at least one valid statement.  

To locate the expected code review comments and answer the RQs in Section~\ref{Research Questions}, we followed the following steps to identify review comments with Python code snippets.

In \textbf{step one}, we filtered out the review comments that are commented by the bot reviewer (i.e., Zuul in OpenStack). Since we aimed at exploring the code snippets in review comments from the perspectives of developers and code reviewers in code reviews, the review comments commented by bot should be excluded. Moreover, we only retained the review comments to Python files because Python files are directly related to Python code snippets. As a result of this step, the number of review comments was reduced to 14,141.

In \textbf{step two}, for answering RQ1, we read through the review comments to identify whether they contain code snippets. We first conducted a pilot labelling with 993 review comments, which were randomly selected among all review comments obtained in \textbf{step one} based on 3\% margin of error and 95\% confidence level~\cite{israel1992dss}, by the first and third authors independently. Conflicts were discussed and resolved with the second author to make sure that they had a consistent understanding of the inclusion criterion of data labelling. We measured the inter-rater reliability and calculated the Cohen’s Kappa coefficient \cite{cohen1960coefficient} as a way to verify the consistency on the labelled comments. The Cohen's Kappa coefficient is 0.86, indicating that the two coders reached a decent agreement. In addition, we also used an automatic detection tool named Guesslang\footnote{\url{https://github.com/yoeo/guesslang}} to help us identify code snippets. It can detect the programming language of a given text by predicting the possibilities of candidate languages. Since we focused on Python code snippets, we used Guesslang to calculate the Python possibility of each review comment so that we could filter out the review comments whose possibilities are lower than a threshold. By comparing the detection results of Guesslang with those of manual labelling during the pilot process, the threshold was set to 0.01. This resulted in a reduction in the number of review comments to 10,790. The first and third authors then labelled the remaining review comments independently.

In \textbf{step three}, for answering RQ2 and RQ3, we extracted the contextual information of identified review comments with code snippets obtained in \textbf{step two}, including the code review discussions and associated source code. To mine the intentions behind the code snippets provided by reviewers, we analyzed the extracted information using the Constant Comparison method \cite{glaser1965constant}. We also checked whether the code snippets were accepted by developers. Specially, we regarded the code snippets as accepted in the following two situations:

\begin{enumerate}[1)]
\item Changes were made in the code by developer(s) based on the code snippets provided by reviewers.

\item Developer(s) clearly showed a positive attitude towards the code snippets provided by reviewers (e.g., \textit{``Thanks for the tip, I'll definitely try that out...\footnote{\url{http://alturl.com/hroeq}}''}).

\end{enumerate}

This data extraction and analysis step was conducted by the first author and the results were verified by two other authors. Conflicts were discussed and addressed by the three authors. The dataset has been provided online for replication purpose \cite{replication-package}. 

\section{Results}\label{Results}
\subsection{Results of RQ1}
\textbf{Motivation:} As an exploratory study on code snippets in code reviews, this RQ aims at investigating the distribution and proportion of review comments that contain code snippets. Such information can help us get a basic view of code snippets in code reviews.

\noindent\textbf{Results:} Table \ref{overview} presents an overview of the review comments and code snippets per project. In general, we obtained a total of 626 review comments that contain code snippets. Compared with the number of all the review comments we analyzed (i.e., 10,790), we can find that code snippets are not prevalently used in code reviews, only account for 5.8\% on average. Concretely, there are 436 review comments containing code snippets that are identified in Nova (out of 7,669, 5.7\%). The proportion of review comments that contain code snippets in Neutron is a little higher than Nova (190 out of 3,121, 6.1\%).

We further investigated how much of review comments having code snippets are provided by code reviewers and developers. From Table \ref{overview}, we can learn that most of the review comments that contain code snippets are provided by reviewers (547 out of 626, 87.4\%). In Neutron, more than 90\% (173 out of 190, 91.1\%) of the code reviews that contain code snippets are provided by reviewers. It is not surprising that most of the code snippets in code reviews are provided by reviewers since reviewers play an important role in the code review process. For example, reviewers usually provide useful suggestions to help developers resolve the issues or improve the design of code, and such suggestions might contain code snippets (Example: ``\textit{Hello, what you think in create a Class to deal with the complexity of calculating the VMs uptime? If you agree, I have a suggestion bellow, something like this: [code snippets]}''\footnote{\url{http://alturl.com/uonk3}}).


\begin{table}[h]
\caption{Overview of Review Comments (RC) and Code Snippets (CS) per Project}
\label{overview}
\begin{tabular}{|l|l|l|l|l|}
\hline
\textbf{Project}  & \textbf{RC}   & \textbf{RC with CS (\%)} & \textbf{Reviewer} & \textbf{Developer} \\ \hline
Nova              & 7669          & 436 (5.7\%)              & 374               & 62        \\ \hline
Neutron           & 3121          & 190 (6.1\%)              & 173               & 17        \\ \hline
\textbf{Total}    & 10790         & 626 (5.8\%)              & 547               & 79        \\ \hline
\end{tabular}
\end{table}


\subsection{Results of RQ2}
\textbf{Motivation:} As shown in the result of RQ1, most (87.4\%) of the code snippets in code reviews are provided by reviewers. However, little is known about the intentions behind the code snippets provided by reviewers. Therefore, this RQ aims at investigating the purposes of provided code snippets. Such information could help us better understand the role of code snippets in code reviews.

\noindent\textbf{Results:} Two high-level categories of purposes of providing code snippets in code reviews are identified (i.e., \textbf{Suggestion} and \textbf{Citation}). Moreover, we further identified six detailed purposes for the aforementioned two categories. To help readers better understand the categories, we provide a review comment example for each purpose. Due to the space limitation, the complete examples can be found in our replication package \cite{replication-package}.

\textbf{1. Suggestion} refers to the situation where reviewers provide code snippets to recommend developers about what they could do or what they should do to improve the quality of code. It contains four detailed purposes, which are presented below:

\begin{enumerate}[1)]
\item \textbf{Improving Code Implementation:} The code snippets are provided to point out alternative solutions or advice to improve the current code in the patchsets (e.g., design or detailed implementation). (Example: ``\textit{I think we could avoid iterating over the keys/creating a new dict/recursion by doing something like: [code snippets]}'')


\item \textbf{Following Code Style:} The code snippets are provided to make the style of the current code consistent with the best code conventions. To improve the readability of the code, reviewers may suggest developers to address various issues, such as indention, spacing, line breaking, bracing, and so on. (Example: ``\textit{style nit: This would (IMO) read way nicer as: [code snippets]}'')

\item \textbf{Correcting Code:} The code snippets are provided to show what kind of mistakes developers have made in current code and to correct the error code. (Example: ``\textit{I think this still isn't correct, unfortunately. You shouldn't translate the variables. This should read e.g.: [code snippets]}'')

\item \textbf{Complementing Code Implementation:} The code snippets are provided to remind developers that the current code implementations are incomplete and they should complement the code with the provided code snippets. (Example: ``\textit{You need (must) implement the OVO compatibility method, to avoid versioning problems: [code snippets]}'')

\end{enumerate}

\textbf{2. Citation} refers to the situation where reviewers cite internal code (e.g., the current code of files in the patchsets) or external code snippets (e.g., the code of other files in the project) in order to supplement necessary information in the review comments. It contains two detailed purposes as presented below:

\begin{enumerate}[1)]
\item \textbf{Elaborating:} The code snippets are cited to help reviewers supplement their explanations or illustrations. (Example: ``\textit{Unrelated observation but this is actually pretty bad pattern when you think about it. If people don't know we do this and to [code snippets], then they will assign the result to the module replacing it globally.}'')

\item \textbf{Providing Context:} The code snippets are cited to provide additional information related to what reviewers had said in the review comments. (Example: ``\textit{and actually these two lines is the difference, compared to the patch on master, where we have: [code snippets]}'')

\end{enumerate}

We also further investigated what are the most common purposes of the code snippets provided by reviewers in code reviews. Table \ref{accept} presents the distribution of each detailed purpose of using code snippets in code reviews. As for the two high-level categories of purposes, we observed that \textbf{Suggestion} is the most common purpose that accounts for 86.5\% (307 + 74 + 52 + 40 = 473 out of 547), which is far more greater than the result of \textbf{Citation} (13.5\%). This finding indicates that code snippets are commonly used by reviewers to convey suggestions to developers. With regards to the detailed purposes, we found that the most common purpose of providing code snippets in code reviews is \textbf{Improving Code Implementation}  (307 out of 547, 56.1\%) whereas \textbf{Providing Context} is the least common purpose (20 out of 547, 3.7\%). 




\subsection{Results of RQ3}
\textbf{Motivation:} As shown in the result of RQ2, the most common purpose of providing code snippets is \textbf{Suggestion}. However, developers could decide whether to accept the suggestions or just ignore them. Therefore, this RQ aims to investigate how much code snippets are accepted by developers. Such information can help to understand the usefulness of the code snippets in code reviews.

\noindent\textbf{Results:} Table \ref{accept} also presents the accept situation for each \textbf{Suggestion} purpose, while ``-'' is used for the two \textbf{Citation} purposes since the accept situation is not applicable for \textbf{Citation} purposes. In general, we notice that in the 473 review comments that contain code snippets with the aim of \textbf{Suggestion} by reviewers, a total of 322 (201 + 56 + 38 + 27 = 322) review comments were accepted by developers, which accounts for 68.1\%. This finding suggests that code snippets can serve as an effective way when providing suggestions in code reviews. 

We further investigated the acceptance of four detailed \textbf{Suggestion} purposes. From Table \ref{accept}, we can learn that the acceptance rates for the four suggestion purposes are similar. \textbf{Following Code Style} is the purpose with the highest rate of being accepted by developers, which corresponds to 75.7\%. The acceptance rate of \textbf{Improving Code Implementation} is the lowest in our cases, which corresponds to 65.5\%. The difference between the highest and lowest rates is only about 10\%. 


\begin{table}[h]
\caption{Distribution and Accept Situation of Code Snippets in Code Reviews for Each Detailed Purpose}
\label{accept}
\begin{tabular}{|l|l|l|l|}
\hline
\textbf{Purpose}                  & \textbf{Num} & \textbf{Acc} & \textbf{Rate} \\ \hline
Improving Code Implementation     & 307            & 201             & 65.5\%        \\ \hline
Following Code Style              & 74             & 56              & 75.7\%        \\ \hline
Elaborating                       & 54             & -               & -             \\ \hline
Correcting Code                   & 52             & 38              & 73.1\%        \\ \hline
Complementing Code Implementation & 40             & 27              & 67.5\%        \\ \hline
Providing Context                 & 20             & -               & -             \\ \hline
\end{tabular}
\end{table}

\section{Implications} \label{Implications}
We conducted a preliminary study to provide a basic view of code snippets on its distribution, purposes, and acceptance in code reviews. Based on our results, we believe that our work could benefit both practitioners and researchers. \textbf{For practitioners}, our work could present constructive suggestions. For example, as indicated by the result of RQ2, \textbf{Following Code Style} is the second most common purpose behind the code snippets provided by reviewers. It implies that developers may lack familiarity with the code conventions in their organization. Necessary education should be conducted to make developers familiar with the code styles adopted in the system. Moreover, tools that can automatically check for compliance with code conventions would be also useful. \textbf{For researchers}, our work highlights promising research directions. For instance, the result of RQ2 provides a taxonomy of six purposes of code snippets used by reviewers. In another aspect, the purposes of code snippets used by developers also needs further exploration. Moreover, as shown in the result of RQ3, 68.1\% of review comments that contain code snippets were accepted by developers, and the reasons why developers did not accept the comments that contain code snippets are worth further investigation.

\section{Threats to Validity} \label{Threats to validity}
We discuss several threats to the validity of this work according to the guidelines proposed by Runeson \textit{et al.} \cite{runeson2009guidelines}, and how these threats were partially mitigated in our study. Internal validity is not considered, since we do not address causal relationships between variables and results.

\textbf{Construct Validity:} In this work, we depended on human activities, which would induce personal bias. To reduce this threat, the review comments were labelled by two researchers. Any disagreements were discussed and addressed with a third researcher. Moreover, we also conducted a pilot labelling to make sure that the two researchers achieved a consensus on what are code snippets. The data extraction and analysis process was conducted by one researcher, and the results were reviewed and checked by other two researchers. These measures can partially alleviate this threat. 

\textbf{External Validity:} Our work considered two most active projects (Nova and Neutron) from the OpenStack community because OpenStack has made a significant investment in code reviews and is widely used in many studies related to code reviews, but we admitted that more projects from diverse communities are needed to increase the generalizability of the study results.

\textbf{Reliability:} All the steps in our study, including the data mining process, manual data labelling, and manual data extraction and analysis, were conducted and discussed by the three authors. Furthermore, the dataset and analysis results of our study have been made publicly available online in order to facilitate other researchers to replicate our study easily \cite{replication-package}. We believe that these measures can partially alleviate this threat.

\section{Conclusions}\label{Conclusions}
This paper reports on a preliminary study that investigated code snippets in code reviews. We chose a widely known community OpenStack, which provides rich code review data. We then collected and labelled 10,790 review comments from two most active projects (i.e., Nova and Neutron) of OpenStack.

According to our results, we notice that code snippets are not prevalently used in code reviews and most of code snippets are provided by reviewers. We also got a taxonomy of the purposes of using code snippets in code reviews with two categories (i.e., Suggestion and Citation) and six subcategories, among which, Improving Code Implementation is the most common purpose. Moreover, among the code snippets with the aim of suggestion, around 68.1\% was accepted by developers.

Our study highlights the role and usefulness of code snippets in code reviews. Given its importance, we plan to extend this work with a large dataset from diverse communities and projects to further investigate code snippets in code reviews (see Section \ref{Implications}).

\balance
\bibliographystyle{ACM-Reference-Format}
\bibliography{ref}

\end{document}